\documentclass[aps,pre,preprint,groupedaddress]{revtex4-1}
\newcommand {\Cv} 		{C_V}

\newcommand  {\rl}		{\rho_\mathrm{L}}
\newcommand  {\rln}		{\rho^{(N)}_\mathrm{L}}
\newcommand  {\rh}		{\rho_\mathrm{H}}
\newcommand  {\rb}		{\rho_\mathrm{b}}

\newcommand  {\rcm}	{r_\mathrm{cm}}
\newcommand  {\rev}	{r_\mathrm{ev}}
\newcommand  {\rhp}	{r_\mathrm{hp}}
\newcommand  {\kB}		{k_\mathrm{B}}
\newcommand  {\Tb}		{T_\mathrm{b}}
\newcommand  {\Tc}		{T_\mathrm{c}}
\newcommand  {\Tbn}	{T^{(N)}_\mathrm{b}}

\newcommand   {\ev}[1]	{\langle#1\rangle}

\usepackage{graphicx}
\usepackage{color}

\begin{document}

%\preprint{}

\title{Finite-size scaling analysis of protein droplet formation}


\author{Daniel Nilsson and Anders Irb\"ack}
%\thanks{}
\affiliation{Computational Biology and Biological Physics,
Department of Astronomy and Theoretical Physics, Lund University,
S\"olvegatan 14A, SE-223 62 Lund, Sweden}


\date{\today}

\begin{abstract}
The formation of biomolecular condensates inside cells often involve intrinsically disordered 
proteins (IDPs), and several of these IDPs are also capable of forming 
droplet-like dense assemblies on their own, through liquid-liquid phase separation. When modeling
thermodynamic phase changes, it is well-known that finite-size scaling analysis can be a valuable tool.
However, to our knowledge, this approach has not been applied before to the computationally 
challenging problem of modeling sequence-dependent biomolecular phase separation. 
Here, we implement finite-size scaling methods to investigate the phase behavior 
of two 10-bead sequences in a continuous hydrophobic/polar protein model. 
Combined with reversible explicit-chain Monte Carlo simulations of these sequences, 
finite-size scaling analysis turns out to be both feasible and rewarding,
despite relying on theoretical results for asymptotically large systems. 
While both sequences form dense clusters at low temperature, 
this analysis shows that only one of them undergoes liquid-liquid phase separation. 
Furthermore, the transition temperature
at which droplet formation sets in, is observed to converge slowly with system size, so that even for our largest 
systems the transition is shifted by about 8\%. Using finite-size scaling analysis, this shift can be estimated and 
corrected for. 
\end{abstract}

%\keywords{}

\maketitle

\section{Introduction}
Advances over the past decade have shown that, in addition to classical membrane-bound 
organelles, various membraneless liquid-like droplets of proteins and nucleic acids 
can be found within living cells~\cite{Brangwynne:09,Banani:17}. The droplets form 
through a liquid-liquid phase separation (LLPS) process, also called coacervation, in which 
intrinsically disordered proteins (IDPs) often play a key role. Furthermore, it has been demonstrated \textit{in vitro} that several 
of these IDPs are able to phase separate on their own~\cite{Nott:15,Molliex:15,Burke:15},
depending on the solution conditions. Phase-separating IDPs can be rich in charged 
residues~\cite{Nott:15}, but can also be dominated by polar and aromatic 
residues~\cite{Burke:15}.   

To rationalize the phase behavior of IDPs and its dependence on solution conditions, a 
variety of theoretical/computational methods have been employed. A widely used 
method is Flory-Huggins mean field theory~\cite{Huggins:41,Flory:42}, and its extension 
to polyelectrolytes  by Voorn and Overbeek~\cite{Overbeek:57}. However, this approach 
is sensitive only to the overall composition of amino acids, but not to their ordering along the 
chains. One way to overcome this short-coming without resorting to explicit-chain simulation 
is offered by the random phase approximation~\cite{Wittmer:93}, which has been
applied to model the phase-separating ability of IDPs with different charge patterns~\cite{Lin:16}.  

By turning to molecular simulation with explicit chains, key approximations made in the 
above approaches can be removed. In addition, structural properties become readily accessible. 
Therefore, despite being computationally costly, recent years have seen a growing number of 
explicit-chain simulation studies of biomolecular 
LLPS~\cite{Dignon:18a,Dignon:18b,Das:18a,Das:18b,Harmon:17,Harmon:18, Qin:16,Robichaud:19}.
In particular, there have been simulations based on coarse-grained lattice or continuous
representations to elucidate sequence determinants of 
phase-separating IDPs~\cite{Dignon:18a,Dignon:18b,Das:18a,Das:18b}.

Another approach, recently applied for the first time to biomolecular LLPS~\cite{McCarty:19,Lin:19}, is 
to rewrite the original polymer problem as a statistical field theory problem that can be investigated 
by field-theory simulation. This approach has the potential to open for studies of system sizes
that are inaccessible with explicit-chain simulation.  

Yet, regardless of whether explicit-chain or field-theory methods are used, the simulated 
systems are finite and as such there is a need to understand how measured properties   
depend on system size. Fortunately, tools for this 
purpose are available, in the form of finite-size scaling theory for droplet formation 
by phase separation~\cite{Binder:80,Biskup:02,Binder:03,Neuhaus:03}. These tools have 
previously been applied to analyze droplet formation in simpler systems such as 
the lattice gas and the Lennard-Jones fluid~\cite{Neuhaus:03,Schrader:09,Zierenberg:15}, but 
we are not aware of any prior study of biomolecular LLPS using these ideas.  

In this paper, we implement finite-size scaling methods to assess the phase 
behavior of two short model proteins, which provide an instructive 
testbed for the analysis methods. While several previous computational 
studies of IDP phase separation focused on the role of charge-charge
interactions, we here consider a hydrophobic/polar (H/P) protein model. One 
of the sequences we study, called A, is alternating (HPHPHPHPHP), 
whereas the other, called B, has a block structure (HHHHHPPPPP).  
Using Monte Carlo (MC) methods, we perform canonical simulations of these 
sequences for a range of system sizes, with up to 640 chains.   
Both sequences form dense multi-chain assemblies surrounded by a dilute 
background at low temperatures, while only small clusters are present 
at high temperatures. However, the sequences differ in phase behavior. 
We show that their phase behavior can be assessed in a systematic fashion 
by finite-size scaling analysis of the simulation data. This analysis 
demonstrates that one of the sequences, A, undergoes LLPS, whereas 
the other, B, does not.

\section{Methods}
\label{sec:methods}

\subsection{Biophysical model}
\label{sec:model}

We study systems consisting of $N$ copies of a polypeptide enclosed in a periodic 
cubic box with volume $V$. The polypeptide is represented as a string of $n$ hydrophobic 
(H) or polar (P) beads.  The length of the bond between two consecutive beads, $b$, is 
kept fixed, while the polar and azimuthal bond angles are both free to vary. In the absence 
of interactions, the bonds have a spherically uniform distribution.  

The beads interact through a pairwise additive potential, $E=\sum_{i<j}E_{ij}$, where the 
sum runs over all intra- and intermolecular pairs of beads in the system. All beads have a 
diameter of $\rev=0.75b$. When two beads $i$ and $j$ are at a distance $r_{ij}<\rev$ 
from each other, the pair potential $E_{ij}$ becomes infinite. Additionally, 
each HH pair interacts through a soft attractive potential with interaction range $\rhp=2b$.  
If $\rev<r_{ij}<\rhp$, the interaction energy is set to $-\epsilon$ (with $\epsilon>0)$. 
Thus, the pair potential can be summarized as       
\begin{equation}
E_{ij}=\left\lbrace
\begin{array}{ll}
\infty, & \mathrm{if}\ r_{ij}<\rev, \\
u_{ij}, & \mathrm{if}\ \rev<r_{ij}<\rhp, \\
0, & \mathrm{if}\ r_{ij}>\rhp,
\end{array}
\right.
\end{equation}
where $u_{ij}=-\epsilon$ when beads $i$ and $j$ are both hydrophobic, and $u_{ij}=0$ otherwise.

Throughout this article, lengths and energies are given in units of $b$ and $\epsilon$, respectively.
 
\subsection{MC simulations}
\label{sec:mc}

We investigate the thermodynamics of droplet formation in this model by using MC methods 
to generate samples from the canonical ($NVT$) ensemble. Of particular interest is the
behavior at the onset of droplet formation. Therefore, given $N$ and $V$, the temperature $T$
is chosen close to the maximum of the heat capacity, by an iterative procedure. 
Simulations at one or several additional temperatures are performed 
when needed to ensure an accurate description of the heat capacity
throughout the transition region.   
The temperature-dependence of the heat capacity is computed by reweighting 
techniques~\cite{Ferrenberg:89}, using data from all simulated temperatures. 

The efficiency of MC simulations depends strongly on the choice of move set. 
We use a set of six elementary moves. Two of the moves update the internal 
structure of individual chains. The first of these is a single-bead move, which
turns a randomly selected non-end bead about the axis through its two 
nearest neighbors. The second one is a pivot-type rotation, where the 
rotation axis goes through a randomly selected non-end bead in a random direction. 
Beads on one side of the selected one are turned as a rigid body about this axis.  

The other four moves are rigid-body translations and rotations 
of either a single chain or a cluster of chains.
In the cluster moves, the clusters are constructed probabilistically using a Swendsen-Wang-type 
algorithm~\cite{Swendsen:87,Irback:13}. The construction is recursive and begins by picking a 
random first cluster member, $i$. Then, each chain $j$ that has an interaction energy $E_{ij}<0$
with chain $i$ is  added to the cluster with probability $p_{ij}=1-\text{e}^{\beta E_{ij}}$, where $\beta=1/\kB T$ 
is inverse temperature ($\kB$ is Boltzmann's constant). This step is iterated until the list of potential 
further cluster members is empty. Finally, the resulting cluster is subject to a trial rigid-body move. 
The form of the statistical weight $p_{ij}$ is such that no Metropolis accept/reject criterion is needed; 
any sterically allowed move is accepted. 

For each choice of $N$, $V$, $T$ and HP sequence, a set of 1--8 trajectories is generated, each 
comprising $10^7$ MC sweeps, where one MC sweep corresponds to $nN$ elementary updates. 
Multiple runs are used for the largest systems, to ensure statistical significance. Statistical 
uncertainties are computed using a jackknife procedure~\cite{Miller:74}. 

\subsection{Finite-size scaling theory}
\label{sec:fss}
  
Droplet formation by phase separation in finite systems is a 
topic that has been extensively studied over the 
years~\cite{Binder:80,Biskup:02,Binder:03,Neuhaus:03}. 
This body of research provides a general framework for finite-size scaling analysis,
which has been tested on systems such as the lattice gas and the 
Lennard-Jones fluid~\cite{Neuhaus:03,Schrader:09,Zierenberg:15}. This section outlines 
some key results that will be used in Sec.~\ref{sec:res}.   

\begin{figure}
\centering
\includegraphics[width=8cm]{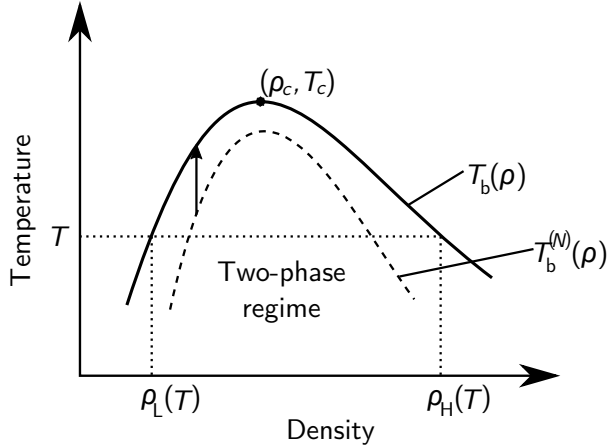}
\caption{
Schematic temperature-density phase diagram of a system that 
undergoes phase separation below an upper critical temperature, $\Tc$, into   
two phases with respectively low ($\rl$) and high ($\rh$) densities. 
In other systems, phase separation may occur 
above a lower critical temperature. Below the so-called binodal curve, $\Tb(\rho)$, the
low and high density phases coexist. At the left branch of the curve, the system transitions 
between the low-density phase and a mixed two-phase regime. In finite systems, 
the transition temperature, $\Tbn(\rho)$, is shifted (dashed line). Finite-size scaling 
theory predicts $\Tbn(\rho)$ to converge toward $\Tb(\rho)$ following the scaling
relation in Eq.~(\ref{eq:shift}) (arrow). 
\label{fig:phasesketch}}
\end{figure}

We consider a $d$-dimensional system of $N$ particles in a volume $V$ at temperatures $T$ 
below an assumed critical temperature $\Tc$. A schematic phase diagram can be found in 
Fig.~\ref{fig:phasesketch}. Under grand canonical conditions, for a given $T<\Tc$ and large 
system size, the system can be in one of two bulk phases with respectively low ($\rl$) and high 
($\rh$) density, depending on the chemical potential. At some value of the chemical potential,  
a first-order transition occurs between these phases. Under canonical conditions, for $T<\Tc$ and 
densities $\rho$ such that $\rl(T)<\rho<\rh(T)$, the system is in a mixed two-phase 
regime, bounded by the binodal curve, $\Tb(\rho)$ (Fig.~\ref{fig:phasesketch}).   
 
Consider now a finite but large system under canonical conditions, 
for a given $T<\Tc$ and $\rho$ just above $\rl(T)$ (Fig.~\ref{fig:phasesketch}). 
At some $\rln(T)>\rl(T)$, the system transitions from a supersaturated dilute
state to a mixed two-phase state. It has been shown that this mixed state consists of 
a single large droplet of the high-density phase in a low-density background, and that  
the linear dimension $R$ of the droplet scales as $R\sim N^{1/(d+1)}$ with 
$N$~\cite{Binder:80,Biskup:02,Binder:03}. This result can be rigorously proven 
for the two-dimensional lattice gas~\cite{Biskup:03}. In brief, the size of the critical 
droplet can be viewed as 
the result of two competing mechanisms for handling a particle excess, $\delta N$. 
One is that the particle excess is absorbed as a density fluctuation in 
the low-density phase, the free-energy cost of which scales as $(\delta N)^2/N$.   
The other mechanism is that a finite fraction of the particle excess forms a dense 
droplet, the free-energy cost of which scales as the surface area of the droplet, 
that is $(\delta N)^{(d-1)/d}$. Assuming 
that droplet formation sets in when these two costs become comparable, one
finds that the linear size of the critical droplet scales as 
$R\sim(\delta N)^{1/d}\sim N^{1/(d+1)}$~\cite{Binder:80,Biskup:02,Binder:03}. 

Using this result, it follows that the finite-size shift of the transition density  
scales as $\rln(T)-\rl(T)\propto N^{-1/(d+1)}$. Correspondingly, 
with $\rho$ rather than $T$ fixed, the transition temperature has a
finite-size shift, given by  
\begin{equation}
\Tbn(\rho)-\Tb(\rho)\propto N^{-1/(d+1)}.
\label{eq:shift}\end{equation} 
Note that this relation implies that the convergence of $\Tbn$ toward its value 
for infinite system size, $\Tb$, is slow.  For comparison, the finite-size shift of a  
regular temperature-driven first-order phase transition scales as $N^{-1}$~\cite{Borgs:92}.        

In finite systems, the transition is not only shifted but also smeared. 
At fixed $\rho$, the smearing, or width, of the transition, $w_T$, may be estimated
as the temperature interval over which $|\beta \Delta F|\lesssim 1$~\cite{Binder:03,Zierenberg:15}, 
where $\Delta F$ is the free-energy difference between the states with and without a droplet.      
Since $\Delta F$ vanishes at $\Tbn$, a Taylor expansion yields
$\beta \Delta F=-[\Delta E/\kB T^2]_{T=\Tbn}(T-\Tbn)$ to leading order. Here, $\Delta E$
is the energy gap, which, assuming that particle interactions are negligible in the 
low-density phase, should scale as the droplet volume, that is 
\begin{equation}
\Delta E\sim N^{d/(d+1)}.
\label{eq:gap}\end{equation} 
It then follows that the smearing of the transition scales as 
\begin{equation}
w_T\propto N^{-d/(d+1)}.
\label{eq:smearing}\end{equation} 
 
When analyzing the droplet transition, a useful property is the specific heat, $\Cv/N$, 
which exhibits a peak at the transition and can be computed from the 
energy fluctuations, using $\Cv=\left(\ev{E^2}-\ev{E}^2\right)/\kB T^2$.    
The transition temperature, $\Tbn$, and smearing, $w_T$, may be defined as, 
the position and width of the specific heat peak, respectively. With increasing $N$, the width of the 
peak, $w_T$, decreases [Eq.~(\ref{eq:smearing})], whereas the height of the peak, 
$C_{V,\max}/N$, increases. With a two-state approximation, one has 
$C_{V,\max}\approx(\Delta E)^2/4\kB T^2$, where $\Delta E$, as before, is the energy 
gap. Using this relation along with Eq.~(\ref{eq:gap}),  one finds that 
\begin{equation}
C_{V,\max}/N\sim N^{(d-1)/(d+1)}.
\label{eq:cvmax}\end{equation}
A slightly different behavior, namely $C_{V,\max}/N\sim N^{d/(d+1)}$, has been 
suggested~\cite{Zierenberg:15}, based on the assumption that $C_{V,\max}/N$ scales inversely proportional
to $w_T$. However, unlike at a regular temperature-driven first-order phase transition, 
in the case of droplet formation, the area under the specific heat peak vanishes in the large-$N$ limit, 
since $\Delta E/N$ does so. Hence, $C_{V,\max}/N$ should grow slower than 
$w_T^{-1}\sim N^{d/(d+1)}$ with $N$, as it does in Eq.~(\ref{eq:cvmax}). 

\section{Results}
\label{sec:res}

\begin{figure}
\centering
\includegraphics[width=8cm]{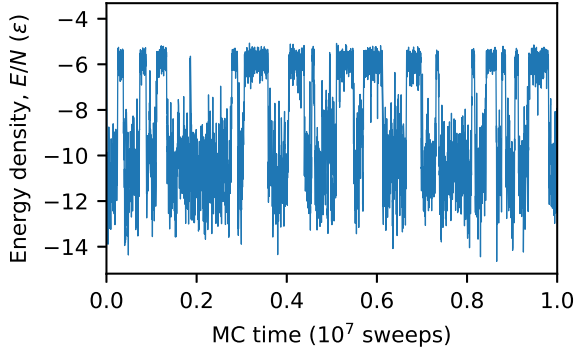}
\caption{
MC evolution of the energy density $E/N$ in a run with $N=640$,   
$T\approx\Tbn$ and $\rb=0.025b^{-3}$, for sequence A. Low and high energies
correspond to states with and without a droplet, respectively. During the 
course of the run, droplet formation and dissolution occur several times.   
\label{fig:rt}}
\end{figure}

Using the model and MC methods described in Sec.~\ref{sec:methods}, we  
conduct thermodynamic simulations of droplet formation by the two sequences 
A (HPHPHPHPHP) and B (HHHHHPPPPP) for a range of system sizes, with up 
to $N=640$ chains. The volume $V$ is adjusted so as to have a given   
bead density $\rb=nN/V$. Most of the calculations are for a bead 
density of $\rb=0.025b^{-3}$, where $b$ is the link length of the chains. For 
comparison, some data for $\rb=0.0125b^{-3}$ and $\rb=0.0375b^{-3}$
are also included. The simulations focus on temperatures near the onset
of droplet formation, and were sufficiently fast for droplets to form and 
dissolve several times during the course of a run, even for the largest 
systems, as illustrated by Fig.~\ref{fig:rt}.

\subsection{Overall characterization}
\label{sec:res_A}

\begin{figure}
\centering
\includegraphics[width=8cm]{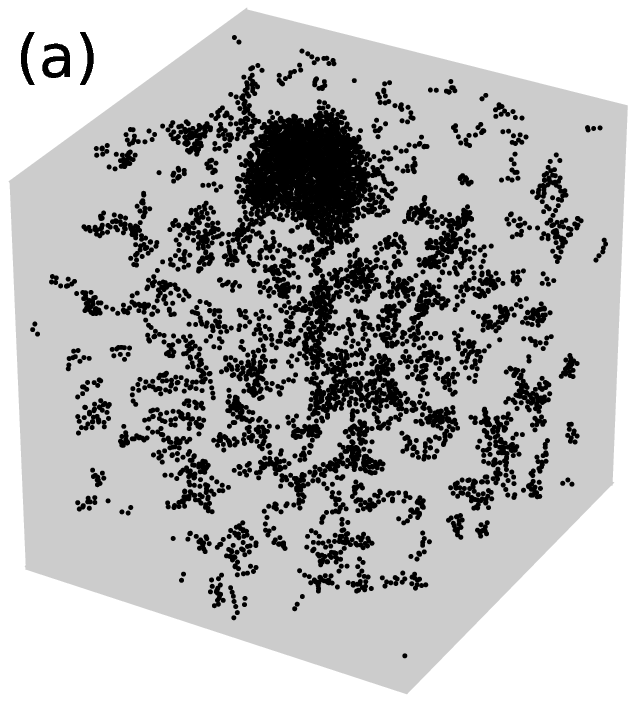}
\includegraphics[width=8cm]{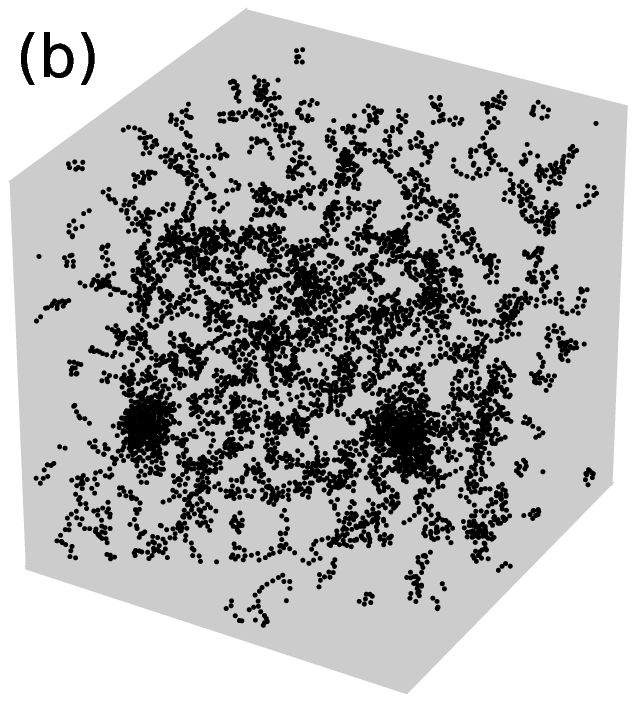}
\caption{
Snapshots showing representative droplet-containing configurations 
from simulations near the temperature at which droplet formation sets in, for 
$N=640$ and  $\rb= 0.025b^{-3}$. %$V=(63.5b)^3$ 
Each bead is shown as a dot. (a) Sequence A, for which a single large droplet 
is observed. (b) Sequence B, which typically forms a few smaller droplets.   
\label{fig:snapshot}}
\end{figure}

At  high temperatures,  the simulated systems are in a disordered state, with only 
small clusters present ($\lesssim 10$ chains). As the temperature is reduced, markedly 
larger clusters, or droplets, appear. Their formation sets in abruptly, in a narrow 
temperature interval, where states both with and without droplets are observed. 
Figure~\ref{fig:snapshot} shows representative snapshots of configurations 
with droplets for both sequences, from simulations with 640 chains. 
For sequence A, a single large droplet can be seen, in a dilute 
background with only small clusters. For sequence B, more 
than one droplet is often present, and the droplets are  
smaller than those formed by sequence A.  A single large 
droplet is what one expects to observe if droplet formation occurs through 
phase separation~\cite{Binder:80,Biskup:02,Binder:03}.   

\begin{figure}
\centering
\includegraphics[width=8cm]{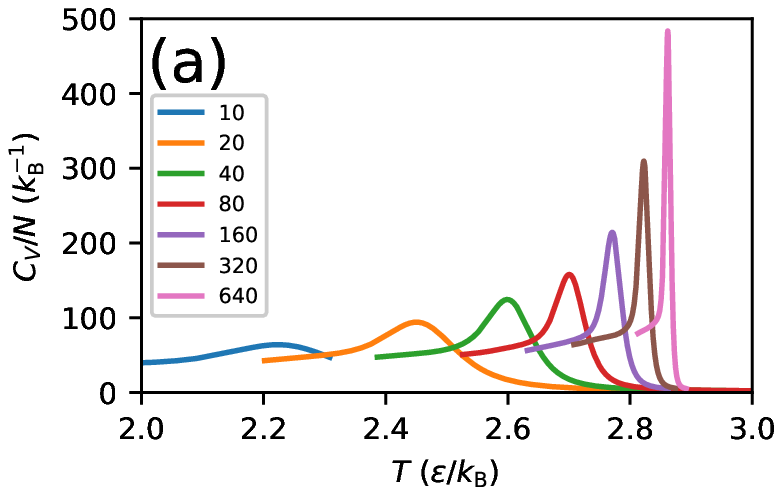}
\includegraphics[width=8cm]{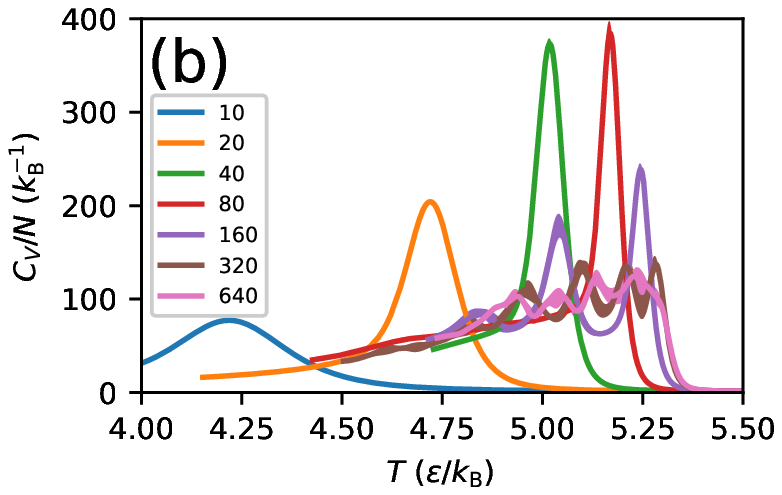}
\caption{
Temperature dependence of the specific heat, $\Cv/N$,  
from simulations with 10--640 chains for fixed $\rb=0.025b^{-3}$.
The curves are computed by reweighting methods~\cite{Ferrenberg:89}, using 
data from canonical MC simulations at several temperatures. 
Shaded bands indicate statistical uncertainties, but are in many cases 
too narrow to be visible. (a) For sequence A, 
the specific heat exhibits a single peak that steadily gets higher  
and narrower with increasing system size. (b) For sequence B, 
the same trend is observed but only for small systems.  In the larger 
systems, sequence B forms more than one droplet, which leads to  
a multimodal specific heat. 
\label{fig:cv}}
\end{figure}

If phase separation occurs, the onset of droplet formation is, furthermore, expected to be   
associated with a divergence in the specific heat (Sec.~\ref{sec:fss}). 
Consistent with this, for sequence A, specific heat data from simulations
with 10--640 chains show a peak that steadily gets higher and narrower 
with increasing system size [Fig.~\ref{fig:cv}(a)]. The corresponding
data for sequence B follow the same trend for small systems [Fig.~\ref{fig:cv}(b)].  
However, for this sequence, at some system size (around 80 chains), the specific heat stops 
growing higher and becomes multimodal. This behavior  
reflects the fact that sequence B forms more than one droplet in the larger 
systems (Fig.~\ref{fig:snapshot}), and implies that this sequence 
does not undergo LLPS.     

\subsection{Finite-size scaling analysis}
\label{sec:res_B}

The above results indicate that, unlike sequence B, sequence A may 
undergo LLPS. To determine whether or not this is the case, we 
next compare simulation data for several quantities with  
predictions from finite-size scaling theory (Sec.~\ref{sec:fss}), focusing 
on sequence A.  

\begin{figure}
\centering
\includegraphics[width=8cm]{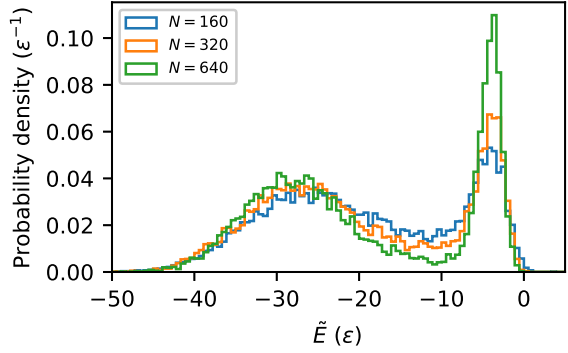}
\caption{
Probability distribution of the shifted and rescaled energy $\tilde E=(E+a)/N^{3/4}$ (with $a=5N\epsilon$)
from simulations with 160, 320 and 640 chains for sequence A, at  
$T\approx\Tbn$ and $\rb=0.025b^{-3}$. Consistent with the predicted scaling relation
$\Delta E\sim N^{3/4}$ [Eq.~(\ref{eq:gap})], the gap between the two peaks 
in $\tilde E$ stays essentially constant, whereas the statistical suppression of intermediate 
energies gets stronger with increasing system size.    
\label{fig:gapscaling}}
\end{figure}

At the onset of droplet formation, due to the coexistence of states with and 
without a droplet, the probability distribution of energy should be bimodal, 
as it is at a regular temperature-driven first-order phase transition. In the 
latter case, the energy gap between the two phases scales linearly with system 
size, corresponding to a non-zero specific latent heat. 
However, at the droplet transition, the energy gap $\Delta E$ should scale as 
the critical droplet volume, or $\Delta E\sim N^{3/4}$ [Eq.~(\ref{eq:gap})].  
Figure~\ref{fig:gapscaling} shows the probability distribution of the shifted and 
rescaled energy $\tilde E=(E+a)/N^{3/4}$, where $a$ is a parameter independent of $E$,
for $T\approx \Tbn$ and $\rb=0.025b^{-3}$, for our three largest systems. 
With larger system size, the probability distribution of $\tilde E$ becomes increasingly 
bimodal in character, due to a stronger suppression of intermediate energies. By contrast,   
the gap between the two peaks in $\tilde E$ stays essentially unchanged, 
in perfect agreement with the predicted scaling of $\Delta E$ [Eq.~(\ref{eq:gap})].

\begin{figure}
\centering
\includegraphics[width=8cm]{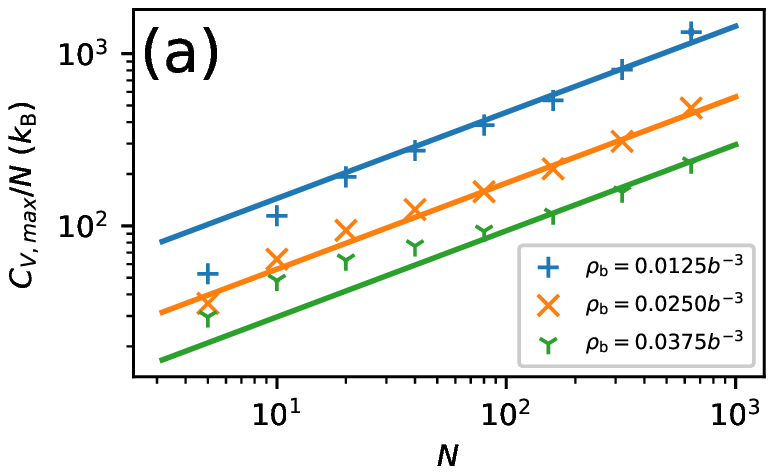}
\includegraphics[width=8cm]{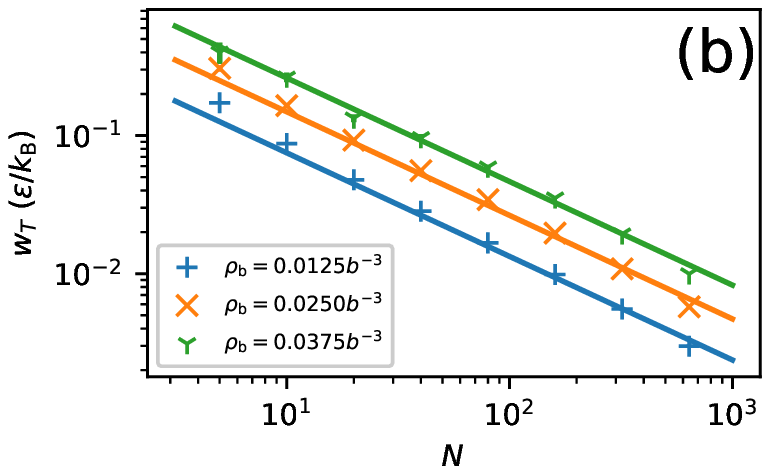}
\includegraphics[width=8cm]{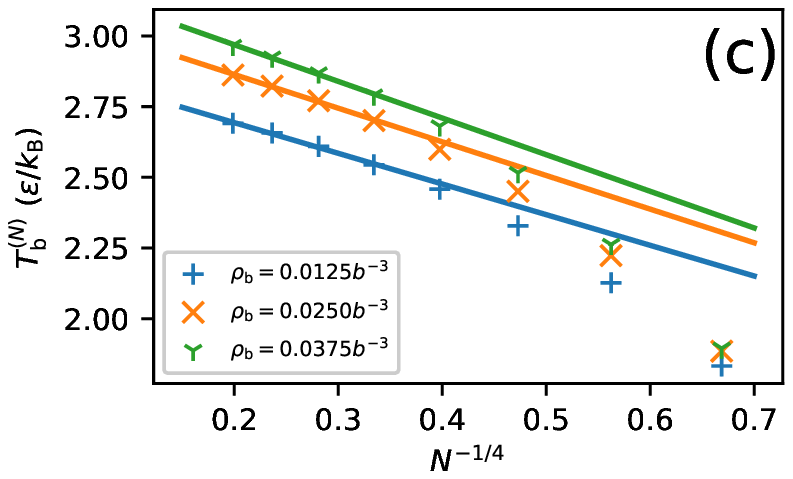}
\caption{
Finite-size scaling analysis at three bead densities $\rb$ ($0.0125b^{-3}$, $0.0250b^{-3}$, $0.0375b^{-3}$)
for sequence A, using data from simulations with 5--640 chains. Lines represent fits of predicted 
scaling expressions from Sec.~\ref{sec:fss} to data for the four largest system sizes.   
(a) Log-log plot of the maximum specific heat, $C_{V,\max}/N$, against $N$. 
The lines are fits of the form $C_{V,\max}/N\sim N^{1/2}$ [Eq.~(\ref{eq:cvmax})].
(b) Log-log plot of the finite-size smearing of the transition, $w_T$, against $N$,  where 
$w_T$ is computed as the length of the temperature interval over which $C_V>0.8C_{V,\max}$. 
The lines are fits of the form $w_T\sim N^{-3/4}$ [Eq.~(\ref{eq:smearing})]. 
(c) The transition temperature $\Tbn$ plotted as a function of $N^{-1/4}$. The lines
are fits of the form $\Tbn=\Tb+cN^{-1/4}$ [Eq.~(\ref{eq:shift})], with $c$ and the 
transition temperature for infinite system size, $\Tb$, as fit parameters. 
The fitted values of $\Tb$ are $\Tb\kB/\epsilon=2.92, 3.10$ and  3.23  for 
$\rb b^3=0.0125, 0.025$ and 0.0375, respectively. 
\label{fig:finsize-A}}
\end{figure}

Assuming this scaling of $\Delta E$ with $N$ [Eq.~(\ref{eq:gap})], the maximum specific 
heat, $C_{V,\max}/N$, should scale as $N^{1/2}$ [Eq.~(\ref{eq:cvmax})].   
Figure~\ref{fig:finsize-A}(a) shows $C_{V,\max}/N$ data against $N$ 
in a log-log plot, for three bead densities $\rb$. Not surprisingly, the data for small systems  
do not follow the predicted scaling relation for large $N$. However, the data 
for the four largest systems ($N=80$--640) match well with the predicted form
for all three bead densities.

Figure~\ref{fig:finsize-A} also illustrates the finite-size smearing and shift of the transition, 
for the same three bead densities. The smearing $w_T$ is expected to scale inversely 
proportional to the energy gap $\Delta E$, or $w_T\sim N^{-3/4}$ [Eq.~(\ref{eq:smearing})].  
From the log-log plot in Fig.~\ref{fig:finsize-A}(b), it can be seen that the data for $w_T$ 
indeed are consistent with the predicted scaling for large $N$.  

The finite-size shift of the transition temperature, $\Tbn-\Tb$, is predicted to scale as
$N^{-1/4}$ [Eq.~(\ref{eq:shift})]. Therefore, Fig.~\ref{fig:finsize-A}(c) shows the data 
for $\Tbn$ plotted against $N^{-1/4}$. As can be seen from this figure, fits of
the form $\Tbn=\Tb+cN^{-1/4}$, with $\Tb$ and $c$ as parameters,
indeed provide a good description of the large-$N$ data ($80\le N\le640$). 
It is worth noting that the scaling of the shift as $N^{-1/4}$, or inversely proportional to 
the linear size of the critical droplet, implies a slow convergence of $\Tbn$ 
toward $\Tb$ with increasing $N$. In fact, for our largest systems 
with 640 chains, $\Tbn$ is still about $8\%$ smaller than the fitted value of $\Tb$. 

To summarize the above analysis, for all properties studied, we find 
that the simulation data for sequence A are consistent with the theoretical 
predictions, which provides strong evidence that this sequence indeed undergoes LLPS.
 
\subsection{Droplet size and structure} 
\label{sec:res_C} 
 
The specific heat data discussed in Secs.~\ref{sec:res_A} and \ref{sec:res_B} 
show that the sequences A and B, despite sharing the same length 
and composition, have different phase behaviors. To understand this 
difference, we next examine some basic structural properties of the droplets 
formed by these sequences. Throughout this section, we focus on data obtained
using $N=640$, $\rb=0.025b^{-3}$ and a temperature near the onset
of droplet formation. 
 
\begin{figure}
\centering
\includegraphics[width=8cm]{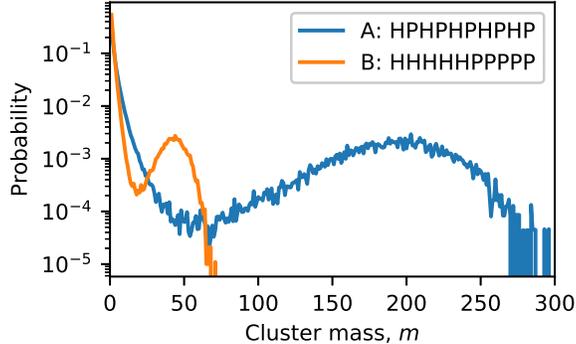}
\caption{
Mass fraction of clusters with $m$ chains, $P(m)$, as obtained using
$N=640$, $\rb= 0.025b^{-3}$ and a temperature near the onset of droplet formation. 
Alternatively expressed, $P(m)$ is the probability that a randomly selected chain 
belongs to a cluster with $m$ chains. Sequence A forms droplets 
containing roughly 200 of the 640 chains, whereas intermediate-mass
clusters are statistically suppressed. 
\label{fig:clusterdist}}
\end{figure}

One important characteristic is the mass of the droplets, or the number
of chains that they contain. It was already noted that sequence A 
forms more massive droplets than sequence B (Fig.~\ref{fig:snapshot}). 
To quantify this assertion, Fig.~\ref{fig:clusterdist} shows cluster 
mass distributions for both sequences. From this figure, it can be seen that, 
in these systems, a typical sequence A droplet accommodates about 200 chains, 
whereas the corresponding number for sequence B is less than 50. 
Also worth noting is the statistical suppression of intermediate-mass 
clusters, which is particularly pronounced for sequence A. If phase 
separation occurs, one expects to observe a single dominant 
droplet~\cite{Binder:80,Biskup:02,Binder:03}, 
as is the case for sequence A.  

\begin{figure}
\centering
\includegraphics[width=8cm]{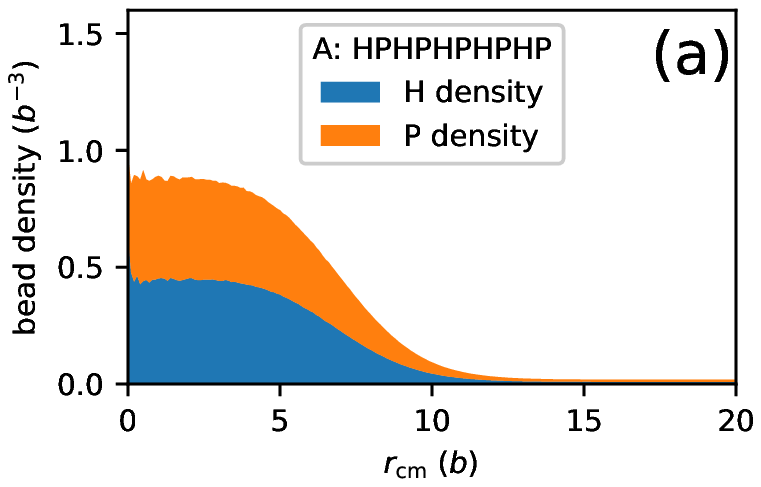}
\includegraphics[width=8cm]{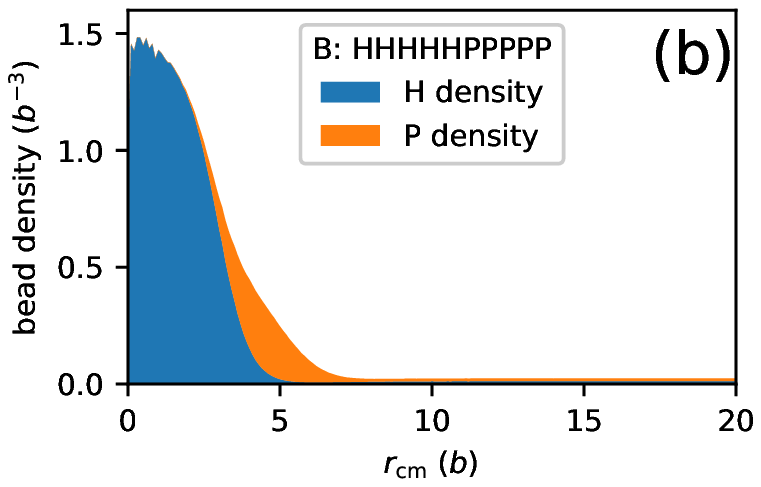}
\caption{
Bead density profiles calculated as a function of the distance $\rcm$ from the center of mass of 
large clusters, for (a) sequence A and (b) sequence B. The data was obtained using 
$N=640$, $\rb= 0.025b^{-3}$ and a temperature near the onset of droplet formation.
A cluster is defined as large if  the number of chains is above a cutoff 
(75 and 20 for sequences A and B, respectively). The total density is split into H and P densities. 
For comparison, a perfect close-packing of the beads would give a total density of $3.35b^{-3}$. 
\label{fig:density}}
\end{figure}

Another basic characteristic is the density of the droplets. Figure~\ref{fig:density} 
shows average bead density profiles around the center of mass of large clusters. 
Here, a given cluster is defined as large if the number of chains exceeds a threshold (75 for sequence A
and 20 for sequence B), and the density is calculated as a function of the distance from   
its center of mass, $\rcm$, counting all beads, whether or not they belong to a 
chain in the cluster. The total density is split into H and P densities. 
The calculated density profiles for sequence A are essentially flat 
at both small and large $\rcm$ [Fig.~\ref{fig:density}(a)], suggesting 
that these densities are representative for the interior of droplets and the 
dilute background, respectively. Using this property, we find that the   
density inside droplets is more than a factor 40 higher than that of the dilute 
background (where the total bead density is $0.019b^{-3}$).  Note also that
the droplets are homogeneous in composition; the H to P ratio is virtually
independent of $\rcm$. 

The droplets formed by sequence B exhibit, by contrast, 
a micellar structure, with a high-density core composed almost 
exclusively of H beads and a corona of mainly P beads [Fig.~\ref{fig:density}(b)]. 
The formation of a hydrophobic core is possible due to the block
structure of this sequence. However, as the sequence is short and contains
a stretch of P beads, this core can only accommodate a small number of chains, which 
explains the low mass of droplets formed by this sequence (Fig.~\ref{fig:clusterdist}). 
The mechanisms of micelle formation by block copolymers have been extensively 
studied by both theory and simulation~\cite{Leibler:83,Nagarajan:89,Milchev:01}. 

While we have seen above that sequence A phase separates, 
it is still not immediately clear whether the dense phase is liquid-like. 
Therefore, we end with a brief assessment of the mobility of 
the chains in droplets formed by this sequence. The analysis uses 
configurations stored at a time interval of $10^3$ MC cycles, which is 
much shorter than the average droplet lifetime of about $2\cdot 10^5$ MC cycles. 
As before, a droplet is a cluster with more than 75 chains.   
We first consider the exchange of chains between droplets and their surroundings. 
To this end, whenever two consecutive snapshots both contain droplets, 
the chain contents of the droplets are compared. Over this
timescale ($10^3$ MC cycles), it turns out that, on average, 44\% of the chains 
present in the original droplet are lost, indicating a fast exchange with 
the surroundings compared to the lifetime of a droplet.

To get a measure of whether also the internal structure of a droplet is dynamic, 
we monitor changes in chain-chain contacts within droplets. To this end, given a
droplet-containing snapshot, we identify all pairs of chains in the droplet that 
are in contact (interaction energy $<$0), and where each chain also 
interacts with at least 15 other chains. The latter condition serves to focus 
the analysis on chain pairs buried in the interior of the droplet. Whenever a 
droplet is present also in the next snapshot ($10^3$ MC cycles later), we 
check the fate of the contacts identified in the first snapshot. On average, 
we find that $54\%$ of the pairs remain in contact, whereas only about 11\% 
are broken due to at least one of the chains leaving the droplet. This leaves 
34\% of the pairs separating due to internal rearrangements of the droplet, 
showing that the internal structure is far from rigid. Thus, the droplets are 
dynamic with respect to both exchange with the surroundings and their 
internal organization. 

\section{Discussion and Conclusions}

It is well-known that finite-size scaling theory provides a powerful tool for analyzing
phase transitions in spin models as well as vapor-to-droplet transitions
in simple liquids. In this manuscript, we have applied these ideas to investigate the
sequence-dependent phase behavior of a simple explicit-chain model for protein droplet formation.  

Of the two specific sequences studied, the block sequence B turned out not to undergo LLPS.
It is worth noting that from data for small systems, one may be led
to the opposite conclusion. In particular, the observed peak in the specific heat is
for small systems 
higher for sequence B than it is for the alternating sequence A, which does 
phase separate. However, above some system size (about 80 chains), 
the maximum specific heat does not increase further for sequence B, in contrast to what is 
observed for sequence A and to what one expects if phase separation takes place. 

For sequence B, we observed micelle formation, rather than the formation of a droplet of
a dense bulk phase. Micelle formation was found to set in at a $kT$ of about 5. Note that the
system need not remain micellar in character well below this temperature. In particular, it is
conceivable that the global free-energy minimum of this system contains bilayer structures at
low temperatures. However, a proper investigation of the low-temperature phase structure
is computationally challenging and beyond the scope of the present article.
   
To determine whether or not sequence A phase separates, simulation data for 
several properties and a range of system sizes were compared with predictions 
from finite-size scaling theory. In this way, the phase behavior can, in principle, 
be investigated in a systematic fashion, but it must be remembered that the theoretical 
results are leading-order predictions for large systems and therefore not necessarily 
valid for the system sizes amenable to simulation. It turned out, however, that 
a scaling behavior consistent with the predicted asymptotic one could be observed
for all properties studied. Hence, taken together, the results of this analysis leave little doubt
that sequence A does indeed phase separate.

It is worth noting that sequences with alternating hydrophobic and polar residues tend
to have a high $\beta$-sheet propensity~\cite{West:99,Hung:17}. The biophysical model 
used in our present calculations cannot describe $\beta$-sheet formation, due to the lack 
of hydrogen bonding. However, it has been shown that droplet formation through LLPS 
sometimes is followed by maturation into a solid-like state containing 
amyloid fibrils~\cite{Patel:15}. In this case, LLPS represents a first step
toward $\beta$-sheet formation.

Among the specific scaling relations studied, the finite-size shift of the 
transition temperature deserves special attention. This shift scales
inversely proportional to the linear size, rather than the volume, of the critical droplet,   
so that $\Tbn-\Tb\sim N^{-1/4}$ [Eq.~(\ref{eq:shift})]. This slow convergence of the transition 
temperature $\Tbn$ toward its value for infinite system size, $\Tb$, makes  
finite-size scaling analysis an important ingredient when determining the 
phase diagram from simulation data. This conclusion is highlighted by the 
magnitude of the relative shift of the transition temperature for sequence A, which 
was found to still be $\sim$8\% for the largest systems with 640 chains.         

Simulation methods, based on explicit-chain or field-theory representations,
offer some distinct advantages over mean-field methods in the study of 
sequence-dependent biomolecular phase separation. However, to exploit
the full potential of the simulations, the system-size dependence of the generated
data needs to be understood and accounted for.  The results presented here 
demonstrate that a systematic analysis of the system-size dependence can 
be both feasible and rewarding.
 
\begin{acknowledgments}
We thank Behruz Bezorg, Sandipan Mohanty and Bo S\"oderberg for stimulating discussions. 
This work was in part supported by the Swedish
Research Council (grant no.~621-2014-4522)
and the Swedish strategic research program eSSENCE.
The simulations were performed on resources
provided by the Swedish National Infrastructure for
Computing (SNIC) at LUNARC, Lund University, Sweden.
\end{acknowledgments}

\end{document}